\documentclass[12pt]{article}

\usepackage{axodraw,graphicx}
\usepackage{amsfonts,amssymb,amsmath}
\usepackage{array,multirow,multicol,amsmath}
\usepackage{mathrsfs,wasysym,ulem,url}

\def\diag{\mathop{\rm diag}}
\def\tr{\mathop{\rm Tr}}
\def\within#1to#2/{#1\mbox{ to }#2}
\def\lbullet#1,#2,#3,#4/{\Vertex(#1,#2)2\Line(#1,#2)(#3,#4)}

\def\ms{\mathscr}
\def\lag{{\ms L}}
\def\sub#1{_{\rm #1}}

\def\bar{\overline}
\def\Eqn#1{Eq.\ (\ref{#1})}
\def\fig#1{Fig.~\ref{#1}}
\def\tabl#1{Table~\ref{#1}}

\title{\bf Involution symmetries and the PMNS matrix\thanks{Talk given by
    PBP at the conference PHENO1 organized by IISER Mohali in April
    2016.}}

\author{\bf 
Palash B. Pal$^1$ and Pritibhajan Byakti$^2$ \\[3mm]
\normalsize $^1$~Saha Institute of Nuclear Physics, Calcutta 700064 \\
\normalsize $^2$~Indian Association for the Cultivation of Science,
Calcutta 700032} 

\date{}

\begin{document}

\maketitle

\begin{abstract}

Lam suggested that the PMNS matrix (or at least some of its elements)
can be predicted by embedding the residual symmetry of the leptonic 
mass terms into a bigger symmetry.  We analyze the possibility that
the residual symmetries consist of involution generators only and
explore how Lam's idea can be implemented.
\\
\noindent\null\hrulefill

\end{abstract}

\section{Introduction}
The quark mixing matrix is very close to the identity matrix.  In
other words, all off-diagonal elements are quite small.  Hence one
might expect that the structure of the quark mixing matrix can be
described by some very small perturbations over the weak basis.  This
is not the case for the lepton mixing matrix, which is called the PMNS
matrix.  Here, some (though not all) of the off-diagonal elements are
quite large, comparable to the diagonal elements.  Therefore some kind
of non-trivial symmetry may be required to understand the structure of
the PMNS matrix.

The PMNS matrix comes out of diagonalization of the mass matrices of
charged leptons and neutrinos.  There are two ways to talk about
symmetries of the mass matrices.  In one way, one starts from a high
energy theory which dictate some symmetries, and sees what part of the
symmetries survive at low energies.  In the other approach, advocated
first by Lam \cite{Lam}, one starts by looking at the symmetries of
the low energy Lagrangian, and tries to see which group can contain
these symmetries.  The bigger symmetry might then determine the PMNS
matrix, or at least some information about its elements.  In this
talk, we are going to discuss some investigations in this second
approach.

\section{PMNS matrix and symmetry generators}\label{s:sg}
The charged current interaction of leptons is governed by the
following term in the Lagrangian:
\begin{eqnarray}
\lag \sub{cc} = \frac{g}{\surd2} \sum_{\ell,\alpha} 
  \bar\ell U_{\ell\alpha} \gamma^\mu L \nu_\alpha W_\mu^+ +
  \mbox{h.c.} \,, 
\end{eqnarray}
and the mass terms are
\begin{eqnarray}
\lag \sub{mass} = - \sum_\ell M_\ell \bar \ell \ell - \frac12
\sum_\alpha m_\alpha \nu_\alpha^\top C \nu_\alpha \,,
\end{eqnarray}
assuming Majorana neutrinos.  Note that the mass terms admit the
following symmetries:
\label{masssym}
\begin{eqnarray}
\nu_\alpha &\to& \eta_\alpha \nu_\alpha \,, \qquad \eta_\alpha = \pm 1; \\*
\ell &\to& \exp(i\varphi_\ell) \ell \,.
\end{eqnarray}
The kinetic terms are also invariant under these transformations.

The symmetry transformations imply that the charged current
interaction is invariant under the change
\begin{eqnarray}
U_{\ell\alpha} \to U_{\ell\alpha} e^{-i\varphi_\ell} \eta_\alpha \,.
\label{Usymm}
\end{eqnarray}
Hence the value of $U_{\ell\alpha}$ is not physical, but the absolute
value is.

According to the basic idea proposed by Lam, we can find
$|U_{\ell\alpha}|$ from the symmetry, without going through the
Lagrangian, if the symmetries of the mass terms are seen as remnant
symmetries after all symmetry breaking.  To implement this idea, we
start with a more general symmetry:
\begin{eqnarray}
\nu \to S \nu \,,  \qquad \qquad
\ell \to T \ell \,,
\label{matsym}
\end{eqnarray}
where $\nu$ and $\ell$, without any index, represent columns of all
three eigenstates of the particles.

In the flavor basis in which the mass terms of the charged leptons are
diagonal, let the neutrino fields be represented by a column
\begin{eqnarray}
\tilde\nu = U \nu \,,
\end{eqnarray}
where this $U$ is the PMNS matrix.  In this basis, the neutrino part
of the symmetry of \Eqn{matsym} would take the form
\begin{eqnarray}
\tilde \nu &\to& U S U^\dagger \tilde\nu \equiv S'
\tilde\nu \,.
\end{eqnarray}

We assume $T$ and $S'$ symmetries are part of a bigger symmetry.  This
symmetry has information about $U$, through
\begin{eqnarray}
S' = U S U^\dagger \,.
\label{S'}
\end{eqnarray}

To proceed, we make some simplifying assumptions.
\begin{enumerate}

\item The symmetry $T$ is discrete.

\item Determinant of all transformations is equal to 1.

\item All generators of the discrete group are involutions, i.e.,
  elements of order 2.

\end{enumerate}
There is a huge body of work in the literature \cite{others} where
only the first two assumptions have been used.  In a recent paper
\cite{P&P} we take the third assumption in addition, and try to find
symmetries consistent with this extra one.  The purpose of this talk
is to summarize essential results of that paper.

The possible $T$ generators, subject to the assumptions taken above,
are as follows:
\begin{eqnarray}
T_e = \diag (1,-1,-1), \qquad 
T_\mu = \diag (-1,1,-1), \qquad 
T_\tau = \diag (-1,-1,1).
\label{T}
\end{eqnarray}
And the possible $S$ generators are:
\begin{eqnarray}
S_1 = \diag (1,-1,-1), \qquad 
S_2 = \diag (-1,1,-1), \qquad 
S_3 = \diag (-1,-1,1) \,,
\label{S}
\end{eqnarray}
with $S'$ defined through \Eqn{S'}.   Note that all the generators
shown here are not independent: there exist the relations
\begin{eqnarray}
T_e T_\mu T_\tau = 1 \,, \qquad S_1 S_2 S_3 =  1 \,.
\label{TTTSSS}
\end{eqnarray}

Let us now define
\begin{eqnarray}
a_{\ell\alpha} = \tr (T_\ell S'_\alpha) = \tr (T_\ell U S_\alpha
U^\dagger) \,.
\label{alalpha}
\end{eqnarray}
It is easily seen, by evaluating the trace, that there is a very
simple relation between these $a_{\ell\alpha}$'s and the modulus of
elements of the PMNS matrix:
\begin{eqnarray}
|U_{\ell\alpha}|^2 = \frac14 \Big(1 + a_{\ell\alpha}\Big) \,.
\label{Ua}
\end{eqnarray}
On the one hand, this relation shows that each $a_{\ell\alpha}$ is
real.  Secondly, it tells us that we can compute $a_{\ell\alpha}$ in
order to find $|U_{\ell\alpha}|^2$. 

How many of these $|U_{\ell\alpha}|^2$ values can be calculated?  It
depends on how many $T$-type and how many $S'$-type generators are
there in our symmetry group.  \Eqn{TTTSSS} tells us that the maximum
number of independent generators of each type is 2.  Thus, there are
only a few possibilities that can arise.  We present these in a
tabular form in \tabl{t:1or2}.
\begin{table}
\caption{Number of generators of the group vs the number of elements
  of the PMNS matrix whose modulus can be determined from
  symmetry. \label{t:1or2}}
\begin{center}
\begin{tabular}{cc|c|c}
\hline
\multicolumn{2}{c|}{No. of generators} & From & Using \\  \cline{1-2} 
$T$-type & $S$-type & \Eqn{Ua} & unitarity \\ 
\hline
1 & 1 & 1 & \\ 
2 & 1 & 2 in a column & full column \\
1 & 2 & 2 in a row & full row \\ 
2 & 2 & 4 in a block & full matrix \\
\hline 
\end{tabular}
\end{center}
\end{table}
Note that unitarity of the PMNS matrix can be used to determine the
modulus of one element if the others in the same row or the same
column are known.  This property has been used in writing the final
column of the table.  Our next task is to examine whether these values
can be consistent with experimental bounds on the PMNS matrix.

Suppose the order of the group element $T_\ell S'_\alpha$ is 
$p_{\ell\alpha}$.  Eigenvalues are of the form $\exp(2\pi i
k/p_{\ell\alpha})$, for integer $k$.  The trace is sum of three such
eigenvalues.  Given that the determinant has been assumed to be 1 and
that trace is real, the only possible solution is of the form
\begin{eqnarray}
a_{\ell\alpha} &=& 1 + \left[ \exp \Big({2\pi i 
  \frac{k_{\ell\alpha}}{p_{\ell\alpha}}} \Big) +  \exp \Big( {-2\pi
  i \frac{k_{\ell\alpha}}{p_{\ell\alpha}}} \Big) \right] \,.
\end{eqnarray}
Then
\begin{eqnarray}
|U_{\ell\alpha}|^2 = \frac12 \left[ 1 + \cos  \Big({2\pi  
  \frac{k_{\ell\alpha}}{p_{\ell\alpha}}} \Big) \right]  =
\cos^2 \Big({\pi  
  \frac{k_{\ell\alpha}}{p_{\ell\alpha}}} \Big)  \,.
\label{cossq}
\end{eqnarray}
For values of $p_{\ell\alpha}$ up to 5, we tabulate the
combinations of $k_{\ell\alpha}$ and $p_{\ell\alpha}$ that give
values which are consistent with experimental bounds.  The result is
shown in \tabl{t:pk}, where we have taken
\begin{eqnarray}
0 \leq k_{\ell\alpha} \leq \frac12 p_{\ell\alpha} \,, \qquad
\gcd(k_{\ell\alpha}, p_{\ell\alpha}) = 1 \,,
\end{eqnarray}
because violating any of these conditions will not give any new value
for $|U_{\ell\alpha}|$.

\begin{table}
\caption{Possible values of the matrix elements consistent with
  experimental limits.\label{t:pk}}
$$
\begin{array}{cccl}
\hline
p_{\ell\alpha}  &
k_{\ell\alpha} &
|U_{\ell\alpha}|^2 & 
\mbox{$|U_{\ell\alpha}|^2$ in range}  \\ 
\hline
\mbox{any} & 0 & 1 & \mbox{none} \\
2  & 1 & 0 &  {\rm none}  \\
3  & 1 & \frac14 & e2, \mu1, \mu2, \tau1, \tau2 \\
4  & 1 & \frac12 & \mu2, \mu3, \tau2, \tau3 \\
5  & 1 & \frac18(3+\surd5) & e1 \\
5  & 2 & \frac18(3-\surd5) & \mu1, \tau1 \\
\hline
\end{array}
$$
\end{table}
%


If the group has more than two generators, there are extra constraints
coming from unitarity.  The reason is the following.  In this case,
there must be more than one generator of one kind --- either
$S'$-type, or $T$-type, or both.  If there are two $S'$ generators,
say $S'_1$ and $S'_2$, then $S'_3$ is also a group element because it
is the product of $S'_1$ and $S'_2$ through \Eqn{TTTSSS}.  Thus, all
three elements of one row (corresponding to the $T$ generator) will
have the form described in \Eqn{cossq}.  Unitarity would require
\begin{eqnarray}
\sum_{\alpha=1,2,3} \cos^2 \Big( \pi  
  {k_{\ell\alpha} \over p_{\ell\alpha}} \Big) = 1 \,.
\label{sumcosS}
\end{eqnarray}
Two $T$-type generators would imply a similar relation for elements of
a column.  Both kinds of equations can be written in the form
\begin{eqnarray}
\cos^2 \Big( \frac{\pi n_1}{N} \Big) + \cos^2 \Big(
  \frac{\pi n_2}{N} \Big) + \cos^2 \Big(\frac{\pi n_3}{N} \Big) = 1 \,.
\label{n/N}
\end{eqnarray}
This is therefore an extra condition that needs to be satisfied, viz.,
we need to find three rational fractional multiples of $\pi$ whose
cosine-squared values would add up to unity.  As it turns out, there
are only two types of solutions to this equation, which are presented
in \tabl{t:n/N}.
\begin{table}
\caption{Solutions of \Eqn{n/N}.  The order of the three number can
  be arbitrary. \label{t:n/N}}

$$
\begin{array}{c@{\quad\quad}l@{\quad\quad}l}
\hline
N & \Big\{n_1, n_2, n_3\Big\} & \mbox{Values of $|U_{\ell\alpha}|^2$} \\
\hline
12 & \Big\{3, 4, 4 \Big\} & \Big\{ \frac12, \frac14, \frac14 \Big\} \\
15 & \Big\{3, 5, 6 \Big\} & \Big\{ \frac18(3+\surd5), \frac14,
\frac18(3-\surd5) \Big\} \\ 
\hline
\end{array}
$$
\end{table}
%

\section{Searching finite Coxeter groups}\label{s:fc}
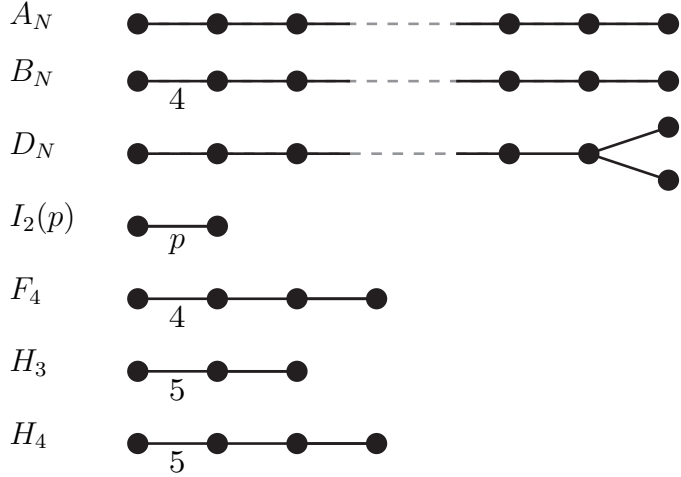
\begin{figure}
\begin{center}
\begin{tabular}{l@{\quad\quad}l}
$A_N$ & \begin{picture}(200,10)
\SetScale{2}
\SetColor{Gray}
\DashLine(0,0)(100,0)2
\SetColor{Black}
\lbullet 0,0,15,0/
\lbullet 15,0,30,0/
\lbullet 30,0,40,0/
\lbullet 70,0,60,0/
\lbullet 85,0,70,0/
\lbullet 100,0,85,0/
  \end{picture}
\\
$B_N$ & \begin{picture}(100,10)
\SetScale{2}
\SetColor{Gray}
\DashLine(0,0)(100,0)2
\SetColor{Black}
\lbullet 0,0,15,0/
\lbullet 15,0,30,0/
\lbullet 30,0,40,0/
\lbullet 70,0,60,0/
\lbullet 85,0,70,0/
\lbullet 100,0,85,0/
\Text(15,-3)[t]{\sc 4}
  \end{picture}
\\[2mm]
$D_N$ & \begin{picture}(100,10)
\SetScale{2}
\SetColor{Gray}
\DashLine(0,0)(70,0)2
\SetColor{Black}
\lbullet 0,0,15,0/
\lbullet 15,0,30,0/
\lbullet 30,0,40,0/
\lbullet 70,0,60,0/
\lbullet 85,0,70,0/
\lbullet 100,5,85,0/
\lbullet 100,-5,85,0/
  \end{picture}
\\[2mm]
$I_2(p)$ & \begin{picture}(20,10)
\SetScale{2}
\SetColor{Black}
\lbullet 0,0,15,0/
\lbullet 15,0,0,0/
\Text(15,-3)[t]{\sc $p$}
  \end{picture}
\\[2mm]
$F_4$ & \begin{picture}(50,10)
\SetScale{2}
\SetColor{Black}
\lbullet 0,0,15,0/
\lbullet 15,0,30,0/
\lbullet 30,0,45,0/
\lbullet 45,0,30,0/
\Text(15,-3)[t]{\sc 4}
  \end{picture}
\\[2mm]
$H_3$ & \begin{picture}(50,10)
\SetScale{2}
\SetColor{Black}
\lbullet 0,0,15,0/
\lbullet 15,0,30,0/
\lbullet 30,0,15,0/
\Text(15,-3)[t]{\sc 5}
  \end{picture}
\\[2mm]
$H_4$ & \begin{picture}(50,10)
\SetScale{2}
\SetColor{Black}
\lbullet 0,0,15,0/
\lbullet 15,0,30,0/
\lbullet 30,0,45,0/
\lbullet 45,0,30,0/
\Text(15,-3)[t]{\sc 5}
  \end{picture}

\end{tabular}
\end{center}

\caption{Coxeter diagrams for finite Coxeter groups.  The notation has
been explained in the text.  We have not drawn groups with more than 4
generators, as well as groups which are products of the groups shown
here.}\label{f:coxeter}
\end{figure}

With this background, let us first see whether we can get information
about the PMNS matrix from finite Coxeter groups.  A Coxeter group is
a group has only involution generators, and is completely determined
by specifying the order of binary products of the generators.  All
finite Coxeter groups are known, and those with 4 or fewer generators
have been shown in \fig{f:coxeter}.

In showing these groups, we have used a graphical notation that is
known as Coxeter diagrams.  Here, each involution generator is
represented by a blob.  The blobs corresponding to two generators
$r_1$ and $r_2$ are not connected by a link if $r_1r_2$ has order 2,
i.e., $(r_1r_2)^2=1$.  If the order of $r_1r_2$ is bigger than 2,
there is a line connecting the two blobs.  If the order of $r_1r_2$ is
larger than 3, the order is written next to the link.  If nothing is
written next to the line, it implies that the order of the product is
3.  Note that we have not shown some groups which do not occur in
chains of arbitrary length and which have more than 4 generators.
Also, note that there are some groups which have been counted more
than once in this figure, viz.,
\begin{eqnarray}
I_2(3) = D_2 = A_2 \,, \qquad I_2(4) = B_2 \,, \qquad D_3 = A_3 \,.
\label{samegroup}
\end{eqnarray}
%

%
\begin{table}
\caption{Acceptable solutions with 3-generator Coxeter groups.  The
  third column shows the value of $N$ given in \tabl{t:n/N}
  which has been used, and the fourth column shows which row (R) or
  column (C) is determined as a result of the exercise.\label{t:3gen}}

$$
\begin{array}{ccccc}
\hline
\mbox{Group} & \mbox{Generators} & N & \mbox{Determined} &
\mbox{Values} \\  \hline \hline
\multirow{8}{*}{$A_3$} & \{S'_1, T_\mu,  S'_2\}
& 12 & R2 & \frac14,\frac14,\frac12 \\
\cline{2-5}
& \{S'_2, T_\mu, S'_1 \}  & 12 & R2 & \frac14,\frac14,\frac12 \\
\cline{2-5}
& \{S'_1, T_\tau,  S'_2\} & 12 & R3 & \frac14,\frac14,\frac12 \\
\cline{2-5}
& \{S'_2, T_\tau, S'_1\} & 12 & R3 & \frac14,\frac14,\frac12 \\
\cline{2-5}
& \{T_e, S'_2, T_\mu \} & 12 & C2 & \frac14,\frac14,\frac12 \\
\cline{2-5}
& \{T_\mu, S'_2, T_e \}  & 12 & C2 & \frac14,\frac14,\frac12 \\
\cline{2-5}
& \{T_e, S'_2, T_\tau \} & 12 & C2 & \frac14,\frac12,\frac14 \\
\cline{2-5}
& \{T_\tau, S'_2, T_e \} & 12 & C2 & \frac14,\frac12,\frac14 \\
\hline
\multirow{8}{*}{$B_3$} & \{T_\mu,  S'_2, T_\tau\}  & 12 &  C2 &
\frac14, \frac12, \frac14  \\
\cline{2-5}
&  \{T_\tau, S'_2, T_\mu\}  & 12 & C2 &
\frac14 , \frac14 , \frac12 \\ 
\cline{2-5} 
& \{S'_3, T_\mu,  S'_1\} & 12 &  R2 
& \frac14 , \frac14 , \frac12 \\ 
\cline{2-5} 
& \{S'_3, T_\mu,  S'_2\}   & 12 & R2 
& \frac14 , \frac14 , \frac12 \\  
\cline{2-5} 
& \{S'_3, T_\tau,  S'_1\}   & 12 & R3 
& \frac14 , \frac14 , \frac12 \\  
\cline{2-5} 
& \{S'_3, T_\tau,  S'_2\}   & 12 & R3 
& \frac14 , \frac14 , \frac12 \\  
\cline{2-5} 
& \{T_\mu,  S'_2, T_e\}   & 12 & C2 
& \frac14, \frac12, \frac14 \\  
\cline{2-5} 
& \{T_\tau,  S'_2, T_e \}   & 12 & C2 
& \frac14 , \frac14 , \frac12 \\  
\hline 
\multirow{4}{*}{$H_3$}& \{T_e,  S'_1, T_\mu\}  & 15 & C1 & 
\frac{3+\surd5}8, \frac14, \frac{3-\surd5}8 \\ 
\cline{2-5} 
& \{T_e,  S'_1, T_\tau\}  & 15 & C1 &
\frac{3+\surd5}8, \frac{3-\surd5}8, \frac14 \\ 
\cline{2-5} 
& \{T_\mu,  S'_1, T_\tau\}  & 15 & C1 &
\frac{3+\surd5}8, \frac{3-\surd5}8, \frac14 \\ 
\cline{2-5} 
& \{T_\tau,  S'_1, T_\mu\}  & 15 & C1 &
\frac{3+\surd5}8, \frac14, \frac{3-\surd5}8 \\ 
\hline
\end{array}
$$
\end{table}
While looking for viable groups from the diagram, we need to follow a
few general guidelines:
\begin{list}{{$\bullet$}}{\itemsep=0pt}

\item Any two $S'$ generators commute.  So two $S'$ generators cannot
  be connected by a line.

\item Same for two $T$ generators.

\item The group must have a 3-dimensional irreducible representation.
  Otherwise the PMNS matrix will turn out to be block-diagonal, with
  zeroes in some off-diagonal places, which is not allowed by
  experimental data.

\end{list}

Consider first groups with 2 generators.  By \Eqn{samegroup}, all
these groups are of the $I_2$ type.  These groups do not have any
3-dimensional irreducible representation, and are therefore not
allowed.


It turns out that there is also no solution for groups with 4
generators.  The reason is the following.  Looking at
\fig{f:coxeter}, we see that if we assign any blob to a $T$-type
generator and the other $T$-type generator to an unlinked blob, we
cannot put the two $S'$-type generators at blobs which will be
connected to both $T$-blobs.  In other words, there will always be at
least one $TS'$ pair that is unconnected by a line.  According to our
notation, that pair will have $p=2$.  But with $p=2$, there is no
acceptable solution, as seen from \tabl{t:pk}.

With 3 generators, however, there are acceptable solutions, and these
have been shown in \tabl{t:3gen}.  As commented earlier, with 3
generators one can predict the absolute values of at most one row or
one column of the PMNS matrix.

\section{Other groups generated by involutions}\label{s:og}
It would be nice if we can have a group that can predict the absolute
values of all elements of the PMNS matrix.  With finite Coxeter
groups, we saw that it is not possible.  Therefore, we now go beyond
these groups.  After all, the machinery developed in Sec.~\ref{s:sg}
applies for any group which can be generated only by involutions, and
can be applied to groups other than finite Coxeter groups.  So we now
search for 4-generator groups outside the realm of finite Coxeter
groups to see whether any group can give us the full PMNS matrix. 
Such groups might even be infinite, in which case we try to see
whether there is a finite subgroup of it that can fit the bill.

We make a search using the GAP repository of finite groups \cite{GAP}.
We start with a presentation with four involution generators, along
with arbitrary integer parameters that denote the order of the
products of various elements, including products of three or more
elements.  Then we take values of these parameters consistent with the
constraints mentioned in \tabl{t:n/N}, and with $k/p$ values given by
the allowed entries of \tabl{t:pk}.  Since only the value of the ratio
$k/p$ is important, we extend the search to ratios of the form
$ka/pa$, with values of $a$ from 1 to 3.  In each case, we tried to
see whether the resulting group has a subgroup with cardinality less
than 3000.  The results are summarized in \tabl{t:og}.  As one sees,
we find only one solution that has 3-dimensional irreducible
representation.

\begin{table}
\caption{Search for solutions outside finite Coxeter
  groups.  The cardinality and the serial Id specifies a finite group
  uniquely in the GAP repository. \label{t:og}}

\begin{center}
\begin{tabular}{cccc}
\hline
Type of & \multicolumn{2}{c}{Finite subgroup} & \multirow2*{3-d
  irrep?} 
\\ \cline{2-3}
 PMNS matrix & Cardinality & Serial Id & \\ 
\hline
\multirow{3}*{12 only} & 12 & 4 & No \\
& 108 & 17 & No \\
& 576 & 8654 & No \\ \hline 
15 only & \multicolumn{3}{c}{(no solution)} 
\\ \hline
12 \& 15 & 1080 & 260 & Yes \\ 
\hline
\end{tabular}
\end{center}

\end{table}

\section{Comments and warnings}
We thus see that it is very difficult to assign involution generators
to flavor symmetries and obtain the entire PMNS matrix as a result.
This statement can be interpreted in two different ways.  In one way,
one can say that the success rate is very small, and the search is
useless.  In another way, one can say that, if we assume that the
involution generators correspond to flavor symmetries, the search can
be narrowed down so that we have only very few alternatives left,
which means that we are close to the end of the tunnel.

In case of a negative outcome, there is also a warning worth
remembering.  If we discard a certain combination of generators from
being the generator of a group that governs the PMNS matrix, it does
not mean that the group itself is discarded.  For a given group, the
generators can be chosen in many ways.  We have discussed only
involution generators.  For any group that we have discarded, it is
perfectly possible that the same group, with some other choice of
generators which are not involutions, gives acceptable entries for the
PMNS matrix.


\begin{thebibliography}{11}
\bibitem{others}
The present conference contribution was a presentation of the
essential results of Ref.~\cite{P&P}.  Detailed citation of earlier
literature can be obtained in that paper.
\end{thebibliography}
\end{document}